\documentclass[journal=jpcafk,manuscript=article]{achemso}
\usepackage[version=3]{mhchem}
\usepackage{titlecaps}
\Addlcwords{the of into}
\usepackage{chemformula} % Formula subscripts using \ch{}
\usepackage[T1]{fontenc} % Use modern font encodings
\usepackage{enumitem}
\usepackage{graphics}
\usepackage{bm}
\usepackage[normalem]{ulem}
\usepackage[symbol]{footmisc}
\Addlcwords{a an the of on at with for and in into from by without}

\usepackage{makecell}
\usepackage{url}
\usepackage{hyperref}

\usepackage{soul}
\usepackage{xr}
\makeatletter
\newcommand*{\addFileDependency}[1]{% argument=file name and extension
 \typeout{(#1)}
 \@addtofilelist{#1}
 \IfFileExists{#1}{}{\typeout{No file #1.}}
}
\makeatother

\newcommand*{\myexternaldocument}[1]{%
    \externaldocument{#1}%
    \addFileDependency{#1.tex}%
    \addFileDependency{#1.aux}%
}
\myexternaldocument{si}

\author{JingChun Wang} \affiliation{Department of Chemistry,
  University of Basel, Klingelbergstrasse 80, CH-4056 Basel,
  Switzerland} \altaffiliation{These authors contributed equally}

\author{Juan Carlos San Vicente Veliz} \affiliation{Department of
  Chemistry, University of Basel, Klingelbergstrasse 80, CH-4056
  Basel, Switzerland} \alsoaffiliation{Present Address: Department of
  Chemistry, Temple University, 1801 N Broad St, Philadelphia, PA
  19122, United States} \altaffiliation{These authors contributed
  equally}

\author{Markus Meuwly}\email{m.meuwly@unibas.ch}
\affiliation{Department of Chemistry, University of Basel,
  Klingelbergstrasse 80, CH-4056 Basel, Switzerland}
\alsoaffiliation{Department of Chemistry, Brown University, RI 02912,
  USA}

\title[]{High-Energy Reaction Dynamics of N$_{3}$}

\begin{document}

\begin{abstract}
The atom-exchange and atomization dissociation dynamics for the
N($^4$S) + N$_2(^1 \Sigma_{\rm g}^+)$ reaction is studied using a
reproducing kernel Hilbert space (RKHS)-based, global potential energy
surface (PES) at the MRCI-F12/aug-cc-pVTZ-F12 level of theory. For the
atom exchange reaction $({\rm N_A N_B} + {\rm N_C} \rightarrow {\rm
  N_A N_C} + {\rm N_B}$), computed thermal rates and their temperature
dependence from quasi-classical trajectory (QCT) simulations agree to
within error bars with the available experiments. Companion QCT
simulations using a recently published CASPT2-based PES confirm these
findings. For the atomization reaction, leading to three N$(^4{\rm
  S})$ atoms, the computed rates from the RKHS-PES overestimate the
experimentally reported rates by one order of magnitude whereas those
from the PIP-PES agree favourably, and the $T-$dependence of both
computations is consistent with experiment. These differences can be
traced back to the different methods and basis sets used. The lifetime
of the metastable N$_3$ molecule is estimated to be $\sim 200$ fs
depending on the initial state of the reactants. Finally, neural
network-based exhaustive state-to-distribution models are presented
using both PESs for the atom exchange reaction. These models will be
instrumental for a broader exploration of the reaction dynamics of
air.
\end{abstract}

\date{\today}

\section{Introduction}
Among the pivotal reactions for high-energy gas flows are the N($^4$S)
+ N$_2(^1 \Sigma_{\rm g}^+)$ atom exchange $({\rm N_A N_B} + {\rm N_C}
\rightarrow {\rm N_A N_C} + {\rm N_B})$ and dissociation reactions to
three N($^4$S) atoms.\cite{park:1990,boyd:2017} The atom exchange
reaction N$_{\rm A}$($^4$S) + N$_{\rm B}$N$_{\rm C}$$(^{1}\Sigma_{\rm
  g}^{+})$ $\rightarrow$ N$_{\rm B}$($^4$S) + N$_{\rm A}$N$_{\rm
  C}$$(^{1}\Sigma_{\rm g}^{+})$ is particularly important for the
energy content and redistribution in reactive high-energy flows as
they occur in atmospheric re-entry (hypersonics) and
combustion. Molecular nitrogen N$_2(^1 \Sigma_{\rm g}^+)$ is also the
main constituent of Earth's atmosphere and abundant in the atmospheres
of Titan, Triton, and Pluto.\cite{atreya:1978,vuitton:2006} Therefore,
the chemistry involving N$_2$ is particularly relevant for
characterizing and understanding the chemical composition and
evolution of such atmospheres. In interstellar environments, electron
irradiation of N$_2$ ices lead to N$_3$ as a product, among
others.\cite{wu:2013}\\

\noindent
The intermolecular interactions and reactive dynamics of N$_3$ have
been studied using experiments and calculations. One of the earlier
computational works considered linear, bent and cyclic-N$_3$ in its
doublet and quartet states from multireference calculations with
particular focus on its unimolecular dissociation.\cite{wodtke:2005}
Specifically, it was reported that bent N$_3$ in its quartet state is
metastable with respect to the N($^4$S) + N$_2(^1 \Sigma_{\rm g}^+)$
ground state products which is also of interest in the present
work. Several potential energy surfaces (PESs) for the quartet state
reaction have been published. One of the earlier attempts was an
empirical London-Eyring-Polanyi-Sato (LEPS) PES.\cite{lagana:1987} A
later PES used CCSD(T) and internally contracted multireference
configuration interaction calculations with different basis sets
(cc-pVQZ, aug-cc-pVTZ) which were fitted to a many-body
expansion.\cite{wang:2003} This was followed by
CCSD(T)/aug-cc-pVTZ-based PESs fitted to different functional
forms\cite{garcia:2008} and a PES for N$_3$ derived\cite{mankodi:2017}
from a PES for N$_4$.\cite{paukku:2013} Most recently, a
permutationally invariant polynomial (PIP) representation for the
N($^4$S) + N$_2(^1 \Sigma_{\rm g}^+)$ PES based on CASPT2/maug-cc-pVQZ
reference energies was reported.\cite{varga:2021}\\

\noindent
Dynamics studies using some of these PESs considered the computation
of thermal rates for the atom
exchange\cite{wang:2003,lago:2006,esposito:2006} and dissociation
reactions.\cite{esposito:1999,mankodi:2017,gu:2023} Comparisons with
experiments were possible for a few high-temperature measurements for
both reactions.\cite{lyon:1972,appleton:1968} For the dissociation
reaction experiments were carried out between 8000 K and 15000
K,\cite{appleton:1968} whereas for the atom exchange reaction
experiments are available at two temperatures (3400 K and at $\sim
1300$ K).\cite{lyon:1972,bar:1967}\\

\noindent
The state-to-state cross sections for atom+diatom reactions are an
essential ingredient for microscopic modeling of reactive gas
flow.\cite{boyd:2017} Using high-level electronic structure
calculations for full-dimensional, reactive and global potential
energy surfaces together with their representation and dynamics
simulations provides - in principle - the necessary information for
their computation.  However, characterizing the state-to-state
reaction cross sections and rates even for triatomics is a daunting
task. Machine-learning-based methods can provide computationally
effective means to exhaustively sample them without the need for
explicit calculation based on, for example, quasi-classical trajectory
(QCT) simulations.\cite{MM.sts:2019,MM.std:2022,MM.std2:2022}\\

\noindent
The present work determines a new full-dimensional reactive PES for
the N($^4$S) + N$_2(^1 \Sigma_{\rm g}^+)$ collision system at the
multi reference configuration interaction (MRCI) level of theory,
complementing the existing PESs for the [NNO], [OON], and [OOC]
reaction systems using the same
methods.\cite{MM.co2:2021,MM.no2:2020,MM.n2o:2020,MM.cno:2018} Given
this PES, thermal rates for atom-exchange and dissociation reactions
are determined together with state-to-distribution (STD) models for
the atom exchange reaction. These are the basis for a comprehensive
treatment of air burning at high temperatures. The results are
compared with experiments, where available, and with QCT simulations
using a recent PIP-represented CASPT2/maug-cc-pVQZ PES which has, to
the best of our knowledge, not yet been used for computing rates for
the atom exchange reaction. For direct comparison, STD models were
also trained from QCT simulations using the CASPT2 PES.\\

\noindent
The remainder of this paper is organized as follows. First the
computational methods are introduced. This is followed by a
characterization of the new PES and its validation for the
N$(^{4}\rm{S})$ + N$_{2}(^{1}\Sigma_{\rm g}^{+})$ atom-exchange and
dissociation reactions. Next, state-to-distribution models are
generated and discussed and finally, conclusions are drawn.\\

\section{Methods}

\subsection{Potential Energy Surface Construction}
The N$_2(^{1}\Sigma_{\rm g}^{+})$ + N$(^{4}\rm{S})$ collision
correlates with the ($^{4}$A$''$) electronic state. For the reference
energies, electronic structure calculations were carried out using the
aug-cc-pVTZ-F12 (AVTZ-F12) basis set at the MRCI-F12 level of theory
using the MOLPRO suite of codes.\cite{MOLPRO} For the MRCI-F12
calculations, multistate CASSCF(15,12) calculations - consistent with
previous work\cite{wodtke:2005,brinck:2002} - were carried out to
correctly characterize the wave function which was then used as the
starting point for the MRCI-F12 calculations. Preference of explicitly
correlated MRCI-F12 over conventional MRCI calculations was given
because a larger number of correctly converged energies across the
grid were obtained (see below) due to the improved basis set
convergence of the MRCI-F12 method.\cite{werner:2011} A total of 8
states, two per spin state (doublet and quartet), and symmetry group
were included in the state-averaged calculations.\\

\noindent
As is well-known for such calculations, there are nuclear geometries
for which either the CASSCF or the MRCI-F12 calculations do not
converge to the correct electronic states or do not converge at
all. Such energies were removed and the grid was reconstructed and
completed (``cleaned'') using a 2D RKHS $V(R,r)$ to evaluate the
missing points. The grid for the reference calculations was $R=[1.4,
  1.6, 1.8\cdots5.0, 5.25, 5.5, 5.8, 6.1, 6.5, 7.0, 8.0, 9.5, 11.0,
  13.0]$ a$_0$, $r=[1.5, 1.6\cdots2.0, 2.05, 2.1, 2.15, 2.25, 2.35,
  2.5\cdots3.1, 3.3\cdots4.1]$ a$_0$, and $\theta=[172.128,
  \\ 161.929, 151.671, 141.399, 131.123, 120.843, 110.563, 100.281,
  90.000]^\circ$ with the other half of the angles defined by
symmetry. Using this cleaned grid the 3-dimensional RKHS
representation was constructed. The permutationally invariant reactive
PES $V({\bf r}) = \sum_{i=1}^3 \omega_i(r_i)V_i({\bf r})$ was
constructed by mixing the three possible channels N$_{\rm A}$N$_{\rm
  B}$+N$_{\rm C}$, N$_{\rm A}$N$_{\rm C}$+N$_{\rm B}$, and N$_{\rm
  B}$N$_{\rm C}$+N$_{\rm A}$ using an exponential switching function
\begin{equation}
    w_{i}(r)=\frac{e^{-(r_i / \rho_j})^{2}} {\sum_{j=1}^{3}
      e^{-(r_j / \rho_j})^{2}}
\end{equation}
for a given structure ${\bf r}$ of N$_3$. The fitted mixing parameters
were $\rho_{1}=\rho_{2}=\rho_{3}=1.0$ a$_0$. The RKHS-PES was constructured from 5481 reference energies, compared
with 7174 energies for the PIP-PES determined at the CASPT2/maug-cc-pVQZ (mAVQZ) level of
theory. The CASPT2 calculations use a smaller CAS(9,9) active space, including the two lowest states in the CASSCF calculations with weights of 0.9 and 0.1, respectively, and an imaginary shift of 0.1 a.u. in the CASPT2 calculations to reproduce the experimentally reported N$_2$ dissociation energy.\cite{varga:2021} \\

\subsection{QCT Simulations}
For thermal rates (atom insertion and full dissociation) $10^7$
independent QCT simulations were carried out for a given
temperature. Semiclassical initial conditions were sampled from
Boltzmann distributions for vibrational quantum number $v$, rotational
quantum number $j$, total angular momentum $J$, impact parameter $b$,
and collision energy $E_{\rm trans}$. 
The thermal rates at given temperature $T$ were then
obtained from
\begin{equation}
 k(T) = g_e(T)\sqrt{\frac{8k_{\rm B}T}{\pi\mu}} \pi b^2_{\rm max}
 \frac{N_{r}}{N_{\rm tot}},
\label{eq:thermal}
\end{equation}
where $g_e(T)$ is the electronic degeneracy factor, $\mu$ is the
reduced mass of the collision system, $k_{\rm B}$ is the Boltzmann
constant, and $N_r$ is the number of reactive trajectories (either
N-atom exchange or N$_3-$atomization). For the exchange reaction,
$g_e(T)=1$ whereas for the full dissociation reaction, $g_e(T) =
q_{N(^4S)}^3/(q_{N(^4S)}q_{N_2}) = 16$. However, to avoid inflating
rates based on state-counts alone, $g_e = 1$ was used for the
atomization reaction as well. The sampling methodology was discussed
in detail in Ref. \cite{MM.cno:2018}. Statistical errors were
quantified through bootstrapping.  For this, $10^6$ samples with 10
random shuffles of the data (100 times of resamples in total) were
used to yield the expectation values of thermal rates along with the
standard deviations.\\

\noindent
For the STD models, QCT simulations were carried out on a grid of
initial conditions. These included $v=[0, 2\cdots8, 10, 12,
  15\cdots30, 34, 38]$, $j=[0, 15, 30\cdots 225]$, and $E_{\rm
  col}=[0.5, 1.0, 1.5\cdots5.0, 6.0, 7.0, 8.0]$ eV. State-specific QCT
simulations were carried out for a total of 80000 trajectories per
initial condition and stratified sampling was used for the impact
parameter $b=[0,b_{\rm max}=12.0\, {\rm a}_0]$. The maximum allowed
propagation time was 75 ps and runs were terminated when the sum of
all atom separations was larger than 12 a$_0$.\\

\subsection{STD training and evaluation}
One approach to creating a model that spans all possible probabilities
for the outcome of a collision between an atom and a diatom is to use
a neural network (NN) representation.\cite{MM.std:2022} Such an NN
consists of seven residual layers with two hidden layers per residual
layer, and uses 11 input and 161 output nodes corresponding to the 11
features representing the initial reactant state and the 161
amplitudes characterizing the product state distributions. For
training, the NN inputs were standardized which ensures that the
distributions of the transformed inputs $x_{i}'$ over the entire
training data are each characterized by ($\bar{x}'_{i}=0$,
$\sigma'_{i}=1$). The NN outputs were normalized which ensures that
the distributions of the transformed outputs $x_{i}'$ over the entire
training data being characterized by ($\bar{x}'_{i}=\bar{x}_{i}$,
$\sigma'_{i}=1$). Standardization of the data generally yields a
faster convergence.\cite{lecun:2012} The loss function was the
root-mean-squared deviation between reference QCT final state
distributions $[P(E_{\rm trans}'), P(v'),P(j')]$ and model
predictions. As the NN outputs are probabilities, being non-negative
even after normalization, a softplus function is used as the
activation function of the output layer.\\

\noindent
For training, the weights and biases of the NN were initialized
according to the Glorot scheme\cite{glorot2010understanding} and
optimized using Adam\cite{kingma2014adam} with an exponentially
decaying learning rate. The NN was trained using TensorFlow
\cite{abadi2016tensorflow} and the weights and biases resulting in the smallest
loss as evaluated on the validation set were subsequently used for
prediction. Overall, final state distributions from 3081 initial
conditions on a grid defined by $(0.5 \leq E'_{\rm c} \leq 5.0)$ eV
with $\Delta E'_{\rm c}=0.5$ eV and $(5.0 \leq E'_{\rm c} \leq 8.0)$
eV with $\Delta E'_{\rm c}=1.0$ eV, $v' =
[0,2,4,6,8,10,12,15,18,21,24,27,30,\\34,38]$, and $0 \leq j' \leq 272$
with step size $\Delta j' = 15$ were generated.\\

\noindent
The results of the QCT simulations starting from 3081 initial
conditions were collected for generating the dataset. Simulations from
1130 initial conditions yield insufficiently low reaction
probabilities $\sum_{v'=0}^{v_{\rm max}'} P(v') < 0.02$ for
convergence and are thus excluded from the training set. Initial
conditions for these low-probability final states are characterized by
low and extremely high $[E_{\rm trans}, v, j]$ values. This is because
for low initial values the atom exchange reaction is improbable to
occur whereas for the largest translational and/or internal energies,
atomization to 3N$(^{4}\rm{S})$ is the dominant final state.\\

\noindent
The final dataset (final state distributions from 2360 initial
conditions) will be referred to ``on-grid'' in the following. A
80:10:10 split of the dataset was randomly drawn for training,
validation and testing.\cite{MM.std:2022} The ``on-grid'' dataset is
to be distinguished from ``off-grid'' final state distributions that
originate from initial conditions which differ in any of the $[E_{\rm
    trans}, v, j]$ quantum numbers used for training. For additional
technical details, see Ref.\cite{MM.std:2022}\\

\section{Results}

\subsection{Topography of the N$_2(^{1}\Sigma_{\rm g}^{+})$ + N$(^{4}\rm{S})$ PES}
The performance of the RKHS-representation for on- and off-grid
reference MRCI-F12 energies is reported in Figure
\ref{sifig:corr-all}. For the single-channel PES the RMSD/MAD for the
training (on-grid) points are 0.03/0.01 kcal/mol across an energy
range of $\sim 10$ eV ($\sim 230$ kcal/mol), see Figure
\ref{sifig:corr-all}A. The permutationally invariant, mixed PES
features RMSD/MAD of 0.45/0.08 kcal/mol with a few outliers at high
energies (Figure \ref{sifig:corr-all}B) whereas evaluating this PES on
off-grid energies leads to RMSD/MAD of 1.50/0.73 kcal/mol, see Figure
\ref{sifig:corr-all}C. Again, small outliers are found at the highest
energies. Restricting the dataset to energies between $-7$ and 0 eV in
Figure \ref{sifig:corr-all}C yields RMSD/MAE of 1.25/0.69 kcal/mol,
respectively. This compares with root mean squared errors ranging from
1.5 to 3.7 kcal/mol for on-grid points covering a comparable energy
range for the PIP-PES.\cite{varga:2021}\\

\begin{figure}
    \centering
    \includegraphics[width=1.0\linewidth]{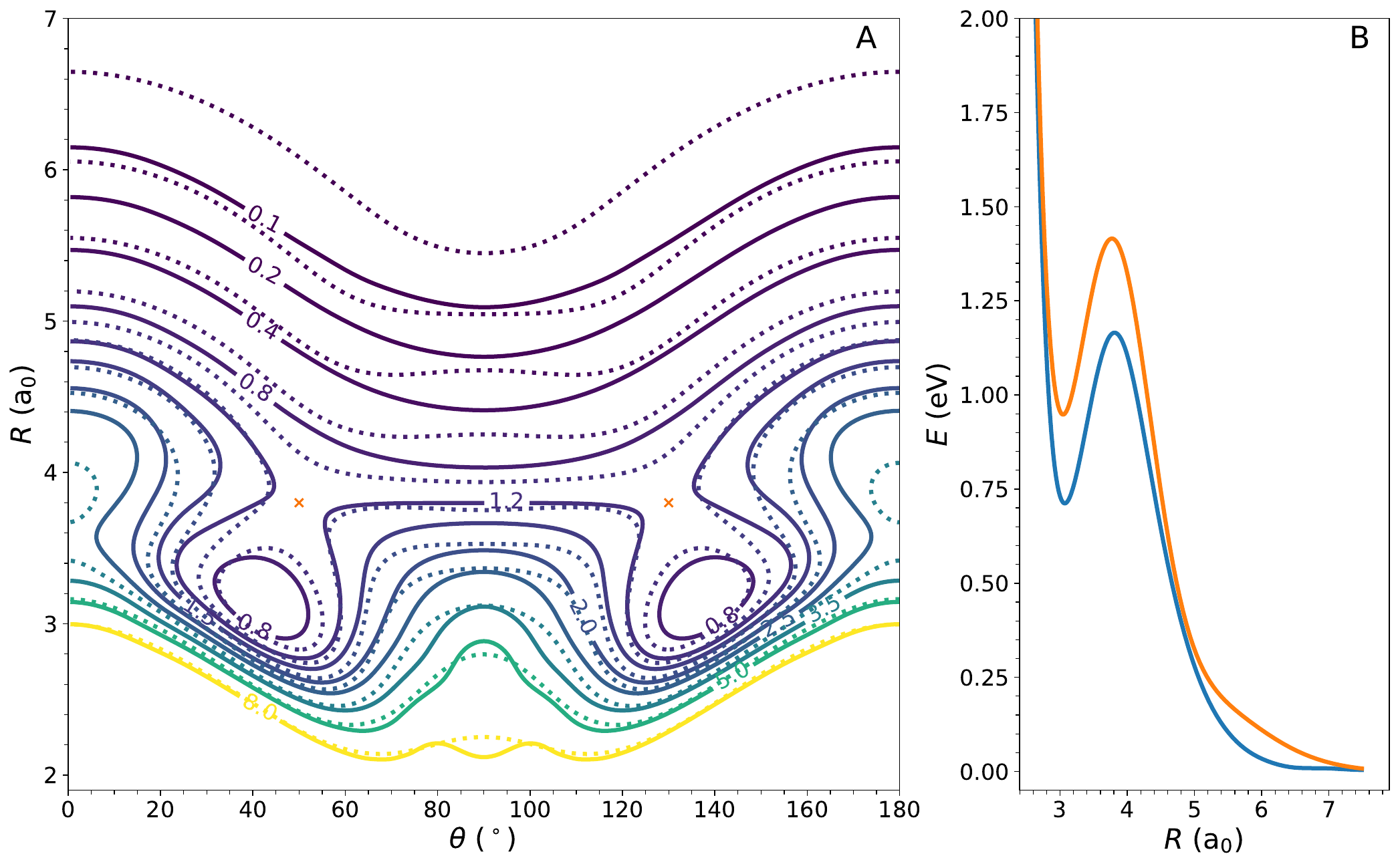}
    \caption{Panel A: 2D contour representations of $V(R,\theta)$ at
      the N$_2$ minimum energy structure for the RKHS representation
      (solid lines, $r_{\rm NN} = 2.39$ a$_0$), and the PIP fit of the
      CASPT2 reference data\cite{varga:2021} (dashed lines, $r_{\rm
        NN} = 2.40$ a$_0$). Isocontours are labelled by energy
      difference in eV relative to $E({\rm N}_2(^1 \Sigma_{\rm g}^{+};
      r=r_e) + E({\rm N})$, of the specific cut of the PES.  Panel B:
      1D cuts $V(R)$ along $\theta = 130^\circ$ in panel A for the
      RKHS-PES (blue) and the PIP-PES (orange).}
    \label{fig:N3_PES_2D-r_eq}
\end{figure}

\noindent
Two-dimensional representations $V(R,\theta; r = r_{\rm eq})$ of the
RKHS-PES (solid lines) and PIP-PES (dashed lines) are reported in
Figure~\ref{fig:N3_PES_2D-r_eq}. The topography of the two PESs is
comparable in their overall shapes, but there are notable differences
in details. Both PESs are symmetric with respect to
$\theta=90^{\circ}$, as expected. The computed minimum energy
structures on the RKHS-PES are clearly identified at $R=3.25$ a$_0$,
$\theta=43.5^{\circ}$ and symmetrically at $R=3.25$ a$_0$,
$\theta=136.5^{\circ}$, which agree closely with positions of the
minima for the PIP-PES, see Figure~\ref{fig:N3_PES_2D-r_eq} and Table
\ref{tab:struc}. The calculated transition state geometries (orange
crosses) are at $R=3.96$ a$_0$, $\theta=52.0^{\circ}$ and
symmetrically at $R=3.96$ a$_0$, $\theta=128.0^{\circ}$, which are
also close to those on the PIP-PES. The geometries and energies for
the critical points of both PESs are summarized in Table
\ref{tab:struc} and compared with earlier calculations as well. Note
that the table reports valence angles $\angle$ $\rm{N_{A}N_{B} N_{C}}$
instead of Jacobi angles $\theta$.\\

\noindent
In the long-range part of the intermolecular interactions the PIP-PES
extends further along $R$ than the RKHS-PES, see the 0.1 eV
isocontours in Figure \ref{fig:N3_PES_2D-r_eq}. The isocontours at
0.2, 0.4 and 0.8 eV for the RKHS-PES point to a conventional
$P_2(\cos(\theta))-$dependence whereas for the PIP-PES the region
between $\theta \in [60,120]^\circ$ is predominantly flat with a faint
undulation superimposed. This is also observed for isocontours at
longer $R-$range and points towards a somewhat different angular
anisotropy of the two PESs. Additional 2D contour plots $V(R,\theta)$
and $V(r_{\rm AB}, r_{\rm BC})$ in Figures \ref{sifig:pes-rtheta} and
\ref{sifig:pes-r1r2} underline the similarities and differences
between the RKHS-PES and PIP-PES.\\

\begin{table}[ht]
\caption{Minimum (MIN) and transition state (TS) geometries for the
  $^{4}$A$''$ electronic state of N$_3$. The critical points were
  calculated using the Nudged Elastic Band
  (NEB)\cite{jonsson:2000,hammer:2016} method. Distances in \AA\/ and
  a$_0$ (in brackets), valence angles in degrees and energies in
  kcal/mol relative to the N$_{2}(^{1}\Sigma_{\rm g}^{+})$ +
  N$(^{4}\rm{S})$ asymptote. $^*$The optimized structures for the
  PIP-based PES were determined using the NEB method to directly
  compare with the results given in the literature.\cite{varga:2021}}
    \begin{center}
    \begin{tabular}{l|l|lll|l|}
    \hline
    & & $r_{e}^{\rm (N_{A} N_{B})}$ & $r_{e}^{\rm (N_{B} N_{C})}$ & $\angle$ $\rm{N_{A}N_{B} N_{C}} $ & $\Delta E$ \\
    \hline
MIN   &                  &                         &                             &               &       \\
      & RKHS MRCI-F12/AVTZ-F12 & 1.26 \AA\/ (2.39 a$_0$) & 1.26 \AA\/ (2.39 a$_0$)      & $118.2^\circ$  & 43.7 \\
      & PIP* CASPT2/mAVQZ & 1.27 \AA\/ (2.40 a$_0$) &  1.27 \AA\/ (2.40 a$_0$)     & $115.5^\circ$  & 53.1* \\
      & PIP CASPT2/mAVQZ  & 1.27 \AA\/              &  1.27 \AA\/                 & $115.6^\circ$  & 53.1\cite{varga:2021} \\
      & CCSD(T)/AVTZ     & 1.27 \AA\/              &  1.27 \AA\/                 & $119.0^\circ$  & 44.7\cite{wang:2003} \\
      & MRCISD(Q)/AVTZ   & 1.27 \AA\/              &  1.27 \AA\/                 & $118.5^\circ$  & 46.2\cite{wodtke:2005}  \\
        \hline
 TS   &                  &                         &                             &               &                        \\
      & RKHS MRCI-F12/AVTZ-F12 & 1.18 \AA\/ (2.23 a$_0$) &  1.51 \AA\/ (2.86 a$_0$)     & $118.5^\circ$  & 47.5 \\
      & PIP* CASPT2/mAVQZ & 1.18 \AA\/ (2.23 a$_0$) & 1.48 \AA\/ (2.80 a$_0$)      & $116.6^\circ$ &  56.5*  \\
      & PIP CASPT2/mAVQZ  & 1.18 \AA\/              & 1.48 \AA\/                   & $116.7^\circ$ &  56.5\cite{varga:2021}  \\
      & CCSD(T)/AVTZ     & 1.18 \AA\/              & 1.50 \AA\/                   & $117.0^\circ$ &  47.1\cite{wang:2003} \\
      & MRCISD(Q)/AVTZ   & 1.18 \AA\/              & 1.50 \AA\/                   & $117.2^\circ$ &  48.7\cite{wodtke:2005}  \\      
        \hline
    \hline
  \end{tabular}
    \end{center}
  \label{tab:struc}
\end{table}

\noindent
The minimum energy and transition state structures and energies for
the present RKHS-PES were determined using the nudged elastic band
(NEB) method.\cite{jonsson:2000} For the local minimum an isosceles
triangle with bond lengths of 2.39 a$_0$ and a valence angle of
$118^\circ$ was found, see Table \ref{tab:struc}. The minimum energy
structure is 43.7 kcal/mol above the dissociation to N$_2$($^1
\Sigma_{\rm g}^+$)+N$(^4{\rm S})$. Using the PIP-PES the bond length
is 2.40 a$_0$ and the valence angle reduces to $115.5^\circ$. This
compares favourably with the minimum energy structure given in the
literature and validates both, the use of the PIP-PES in the present
work and the NEB approach.\cite{varga:2021} The minimum energy
structure for the PIP-PES is, however, 53.1 kcal/mol above the
N$_2$($^1 \Sigma_{\rm g}^+$)+N$(^4{\rm S})$ asymptote, see Table
\ref{tab:struc}. For comparison, two additional minimum energy
structures at the CCSD(T) and MR-CISD+Q levels of
theory\cite{wang:2003,wodtke:2005} using the aug-cc-pVTZ basis sets
were included in Table \ref{tab:struc}. Both confirm the results from
the present RKHS-PES based on MRCI-F12/AVTZ-F12 calculation. Hence,
the differences in the stabilization energy of triangular N$_3$
relative to the N$_2$($^1 \Sigma_{\rm g}^+$)+N$(^4{\rm S})$ limit are
mainly due to the methods and basis sets used.\\

\noindent
To verify this, the dissociation energy of N$_2 (^1 \Sigma_{\rm g}^+)$
was determined using the AVTZ-F12 and AVQZ-F12 basis sets for which
$D_{\rm e}^{\rm AVTZ-F12} = 222.0$ kcal/mol and $D_{\rm e}^{\rm
  AVQZ-F12} = 222.6$ kcal/mol were obtained, compared with $D_{\rm
  e}^{\rm mAVQZ} = 228.7$ kcal/mol at the CASPT2/mAVQZ
level\cite{varga:2021} which differs by 6 to 7 kcal/mol from the
MRCI-F12 calculations. Hence, part of the difference $\Delta E \sim
10$ kcal/mol for the N$_3$ minimum energy relative to N$_2$($^1
\Sigma_{\rm g}^+$)+N$(^4{\rm S})$ (see Table \ref{tab:struc}) is
explained by differences in the stabilization energy of N$_2$ when
using MRCI-F12 or CASPT2 levels of theory.\\

\noindent
For the transition state structure the findings are comparable. The
geometries of the TS using the NEB method for the RKHS-PES and PIP-PES
compare favourably whereas the energy relative to the asymptote
differs by 9.0 kcal/mol, see Table \ref{tab:struc}. The TS structure
and energy from the literature\cite{varga:2021} are confirmed by the
NEB calculations using the PIP-PES. Again, using the smaller AVTZ
basis set confirms both, the structure and energy of the TS relative
to the N$_2$($^1 \Sigma_{\rm g}^+$)+N$(^4{\rm S})$ limit from the
RKHS-PES. Notably, the height of the barrier between MIN and TS
structures is 3.8 kcal/mol for the RKHS-PES compared with 3.4 kcal/mol
for the PIP-PES and 2.4 kcal/mol and 2.5 kcal/mol from the two earlier
calculations using the aug-cc-pVTZ basis
set.\cite{wang:2003,wodtke:2005} Notably, the two PESs considered in
the present work are based on different methods (MRCI-F12 vs. CASPT2)
and two different basis sets (AVTZ-F12 vs. mAVQZ). Given these
differences, the rather close agreement for the geometries and
relative energetics is notable. A final comparison between MRCI+Q/AVTZ
(used in earlier work) and MRCI-F12/AVTZ-F12 levels was carried out
for a 1d-cut with $r_{\rm NN} = 2.3$ a$_0$, $\theta=50^\circ$, and $R
\in [2.4, 8.0]$ a$_0$. The stabilization energies of N$_3$ towards
dissociation to N$_2$ + N are 18.3 kcal/mol and 17.7 kcal/mol,
separated by barrier heights of 27.3 kcal/mol and 28.7 kcal/mol,
respectively. Again, the effect of the basis set on relative energies
is of the order of $\sim 1$ kcal/mol.\\

\subsection{Validation of the PES}
\begin{figure}
    \centering
    \includegraphics[width=0.9\linewidth]{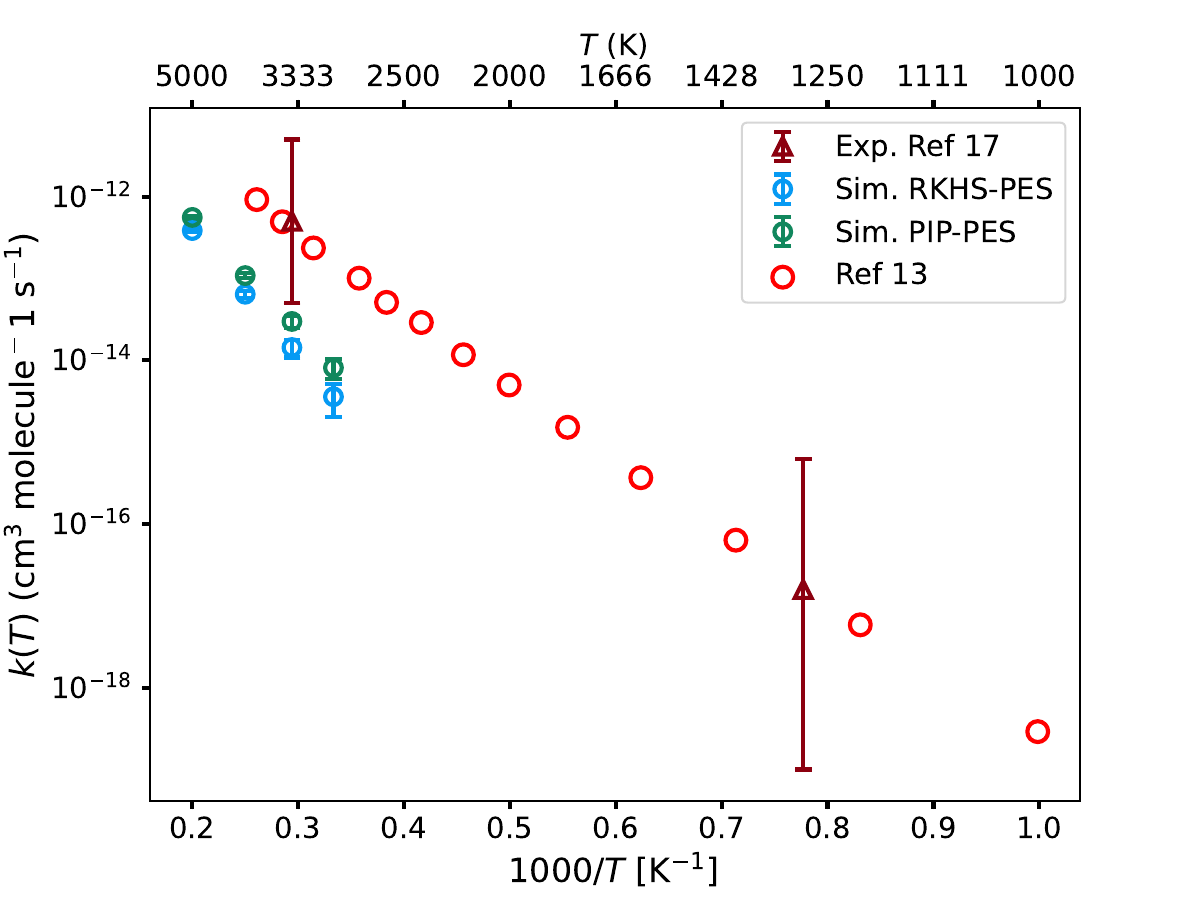}
    \caption{Thermal rates for the atom-exchange reaction from QCT
      simulations on the present RKHS-PES (blue) compared with those
      on the PIP-PES (green) computed from $10^7$ trajectories for
      each temperature considered, experiment (red
      triangles)\cite{lyon:1972} with error bars from a theoretical
      study\cite{wang:2003} and from transition state theory (red
      circles).\cite{lago:2006} The number/percentage of reactive
      trajectories from simulations using the RKHS-PES at 3000, 3400,
      4000, and 5000 K range from 58 ($\sim 10^{-4}$\%) to 4192 ($\sim
      4 \times 10^{-2}$\%), compared with 124 ($\sim 10^{-3}$\%) to
      6342 ($\sim 6 \times 10^{-2}$\%) for the PIP-PES.}
    \label{fig:exch}
\end{figure}

\begin{figure}
    \centering
    \includegraphics[width=0.9\linewidth]{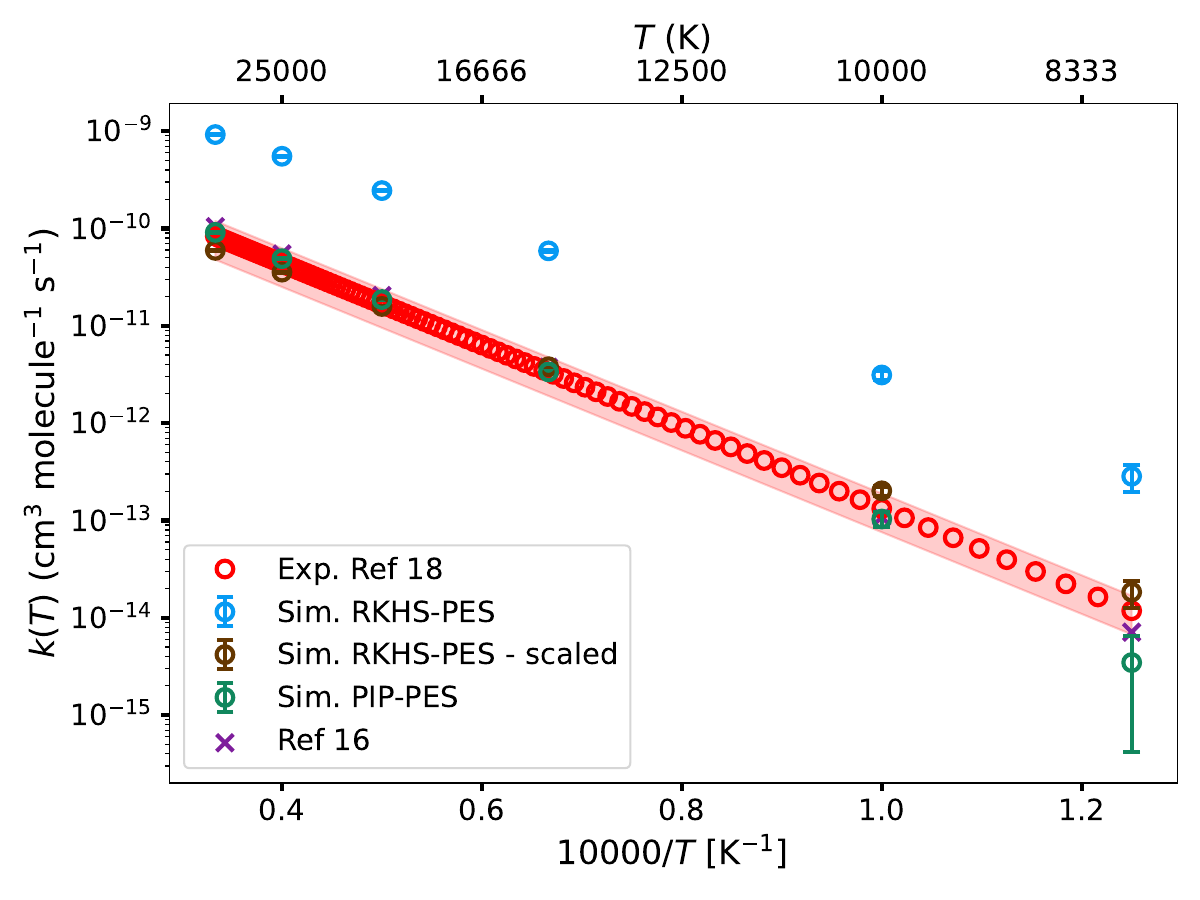}
    \caption{Thermal rates for the atomization reaction from QCT
      simulations on the present RKHS-PES (blue) and the PIP-PES
      (green) computed from $10^7$ trajectories for every temperature
      considered. The experimental data are the red
      circles.\cite{appleton:1968} For both PESs, statistical errors
      are reported from bootstrapping using 10 sub-batches with a
      $10^6$ sample size and conducted 10 times in random order. For
      $T = 8000$, 10000, 15000, 20000, 25000, and 30000 K, the number
      of reactive trajectories ($10^7$ for each $T$) on the RKHS-PES
      range from $\sim 10^{-3}$ \% to $\sim 8$ \% (435, 3899, 64713,
      239998, 489896, 759432) compared with $\sim 10^{-4}$ \% to $\sim
      2$ \% (17, 406, 11803, 57065, 136537, 235190) for the PIP-PES.
      Hence, for converged rates at 8000 K a larger number of
      trajectories would be required. To demonstrate the correct
      $T-$dependence of the rates from the RKHS-PES, the results are
      scaled (brown) to best overlap with the experimental
      reference. The purple crosses are from QCT calculations on the
      PIP-PES\cite{gu:2023} which agree closely with the present
      simulations (green).}
      \label{fig:diss}
\end{figure}

To validate the present RKHS-PES, thermal rates for the atom-exchange
and dissociation reactions were determined, see Figures \ref{fig:exch}
and \ref{fig:diss}. First, it is of interest to briefly consider the
convergence with respect to the number of reactive trajectories for a
given state-to-state transition $(v = 21, j = 30, E=2.0) \rightarrow
(v' = 20, j' = 34)$. With both PESs, $10^7$ QCT trajectories were run
and the rates were determined using $10^5$, $10^6$, and $10^7$
trajectories. The ratios $N_{\rm reactive}/N_{\rm tot}$ for the
RKHS-PES are 0.00558, 0.00550, and 0.00553 compared with 0.00163,
0.00180, and 0.00177 for the PIP-PES, respectively. Hence, convergence
for state-to-state transitions is typically achieved for $\sim 10^5$
trajectories starting from a given initial condition.\\

\noindent
{\it Exchange Reaction:} The results for the atom exchange reaction
(N$_{\rm A}$N$_{\rm B}$+N$_{\rm C}$ $\rightarrow$ N$_{\rm A}$N$_{\rm
  C}$+N$_{\rm B}$) from simulations at 5000, 4000, 3400, and 3000 K
together with associated error bars from bootstrapping using the two
reactive PESs are given in Figure \ref{fig:exch}. The electronic
degeneracy factor for this reaction is $g_{\rm e} = 1$. The fraction
of reactive trajectories ranges from $\sim 10^{-2} \%$ to $\sim
10^{-4} \%$. All computed rates are smaller than the TST rates (red
open circles). Compared with the only experiment available in this
temperature range (at 3400 K), the computed $k(T)$ is within error
bars from the two QCT-derived rates, but with somewhat better
agreement for the PIP-PES. Since the fraction of reactive trajectories
is only $\sim 10^{-4}$ \% at 3000 K, it is unfeasible to determine
thermal rates directly from QCT simulations as the estimated number of
trajectories required for convergence would be on the order of
$>10^9$. The overall temperature-dependence is correctly described by
the present simulations although the slope of $k(T)$ appears slightly
steeper than in the experiments.\\

\noindent
{\it Atomization Reaction:} For the atomization reaction N$_2$($^1
\Sigma_{\rm g}^+$)+N$(^4{\rm S})$ $\rightarrow$ N$(^4{\rm
  S})$+N$(^4{\rm S})$+N$(^4{\rm S})$, the thermal rates are reported
in Figure \ref{fig:diss}. Formally, the degeneracy factor based on
statistical mechanics is $g_{\rm e} = 16$; however $g_{\rm e} = 1$ was
used in the following to not inflate rates based on symmetry
considerations only. The temperature dependence of both QCT-based
simulations (blue for RKHS-PES and green for PIP-PES) follows the
experimentally reported rates (red).\cite{appleton:1968} The computed
$k^{\rm RKHS}(T)$ from simulations using the RKHS-PES is larger by
about one order of magnitude than the experiment (within error
bars). Simulations using the PIP-PES yield $k^{\rm PIP}(T)$ that
follows the experimentally reported rates rather
closely. Nevertheless, $k^{\rm RKHS}(T)$ still correctly describes the
reported $T-$dependence if scaled with a constant factor (brown
symbols in Figure \ref{fig:diss}).\\

\noindent
The PIP-PES was recently used in simulations and a NN-based model for
the atomization reaction.\cite{gu:2023} For equal rotational and
vibrational temperatures these results can be directly compared with
those obtained in the present work, see magenta crosses and green
circles in Figure \ref{fig:diss}. The close agreement is a mutual and
independent validation of both QCT-simulations.\\

\noindent
Overall, the RKHS- and PIP-based PESs yield comparable results
with differences in the details. QCT simulations for the atom exchange
and atomization reactions yield thermal rates $k(T)$ in good to
reasonable agreement with experiments within error bars, whereas for
the atomization reaction, the PIP-PES performs better but the results
using the RKHS-PES also show the correct $T-$dependence. The
possible reason for the larger $k(T)$ of the atomization reaction for
the RKHS-PES is the lower barrier to reach the transition state coming
from N$_2$($^1 \Sigma_{\rm g}^+$)+N$(^4{\rm S})$, see Table
\ref{tab:struc}, and the lower N$_2-$dissociation energy.\\

\subsection{Final State Distributions and NN-Based Model}
Next, the final state distributions for translational, vibrational and
rotational degrees of freedom were determined for the atom-exchange
reaction using the two PESs. As reported above, both PESs (RKHS-PES
and PIP-PES) perform comparably for this reaction. The product state
distributions $P(E_{\rm trans}')$, $P(E_{\rm int}')$, $P(v')$, and
$P(j')$ from particular initial conditions $[E_{\rm trans}, v, j]$ are
analyzed and compared in the following and STD models were generated
for both PESs.\\

\begin{figure}
    \centering \includegraphics[width=\linewidth]{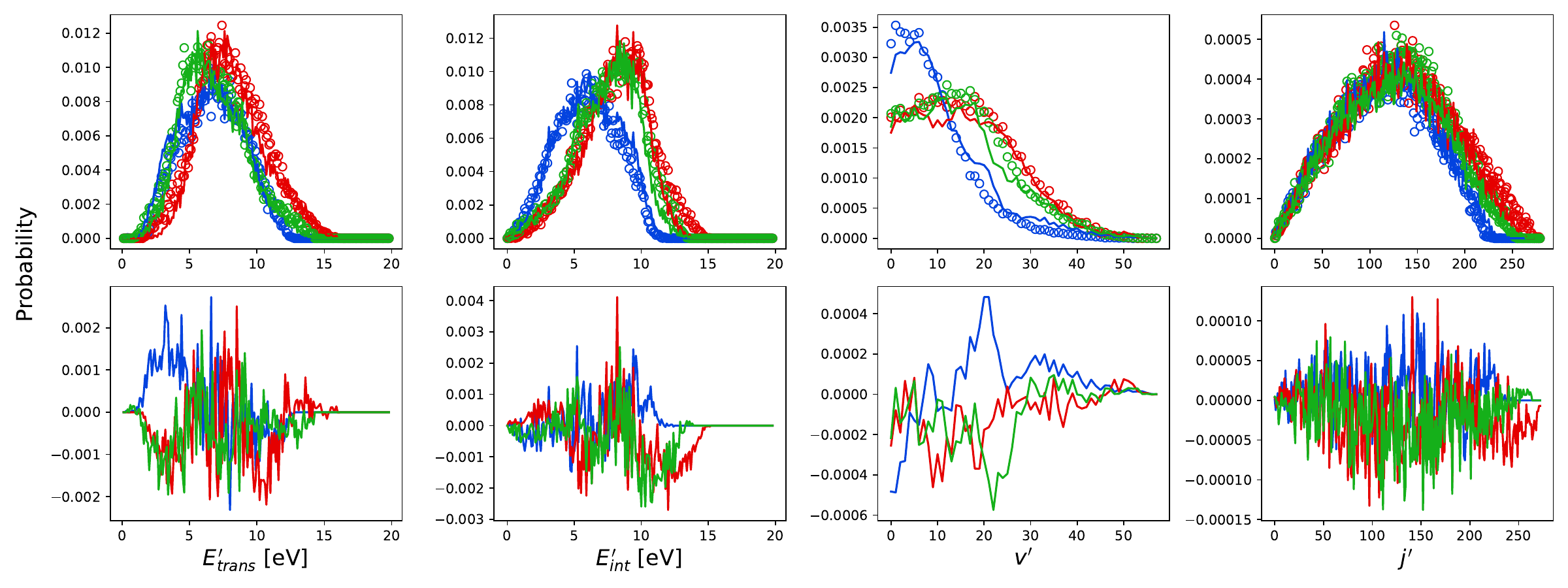}
    \caption{Product QCT distribution for the $^{4}$A$''$ PES$_{\rm
        Truhlar}$ and PES$_{\rm Unibas}$ from 3 initial conditions:
      [$v=2,j=180, E_{\rm trans}=5.0$ eV] (blue), [$v=6,j=225, E_{\rm
          trans}=4.5$ eV] (red) and [$v=10,j=180, E_{\rm trans}=5.0$
        eV] (green). The final $E_{\rm trans}'$, $E_{\rm int}'$, $v'$,
      and $j'$ are plotted as a function of the reaction
      probability. The probability is computed using histogram
      binning. The top panels show the final state distributions of
      the two PES's, RKHS-PES (solid line) and PIP-PES (open circles),
      and the bottom panels report the difference between the
      distributions of the two PESs in the corresponding colors.}
    \label{fig:dist-detail2}
\end{figure}

\noindent
It is of interest to compare final state distributions from QCT
simulations using both PESs for a few representative but randomly
chosen initial conditions, see Figure \ref{fig:dist-detail2}. Results
for three initial conditions using the RKHS-PES as detailed in the
caption of Figure \ref{fig:dist-detail2} are shown as solid lines and
compared to those for the PIP-PES (open circles). The top row in
Figure \ref{fig:dist-detail2} reports the actual final state
distributions whereas the bottom row shows the difference between the
distributions using the two PESs. The overall shapes for the final
state distributions for the same initial state using the two different
PESs are comparable for these particular initial conditions,
specifically for $P(E_{\rm trans}')$, $P(E_{\rm int}')$, and $P(j')$,
whereas for $P(v')$ slight differences are present.\\

\begin{figure}
    \centering \includegraphics[width=\linewidth]{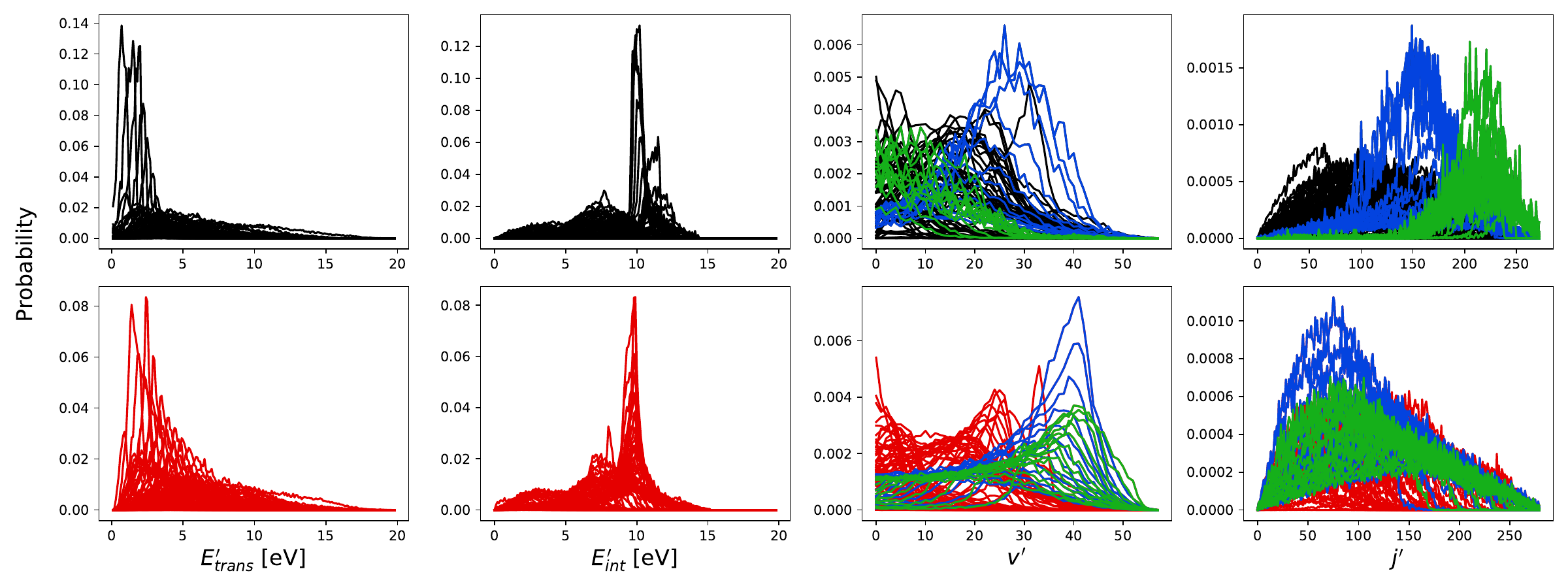}
    \caption{Product QCT distribution for the $^{4}$A$''$ RKHS-PES and
      PIP-PES for several sets of initial conditions.  The final
      $E_{\rm trans}'$, $E_{\rm int}'$, $v'$, and $j'$ are plotted as
      a function of the reaction probability.  The upper row contains
      final distributions for the RKHS-PES, while the bottom row
      contains final distributions for the PIP-PES using the same sets
      of initial conditions. Blue and green curves in both panels
      correspond to initial conditions $(v=30, j=150)$ and $(v=21,
      j=195)$, respectively, including all $E_{\rm trans}$.}
    \label{fig:dist-all-jw}
\end{figure}

\noindent
Next, a broader set of initial conditions was generated. Specifically,
all 4 combinations of small and large values for initial $v$ and $j$
for a range of initial translational energies $E_{\rm trans}$ and
final state distributions were determined from QCT simulations using
both PESs, see Figure \ref{fig:dist-all-jw}. Qualitatively, $P(E_{\rm
  trans}')$, $P(E_{\rm int}')$ and $P(v')$ from simulations using the
two PESs agree with each other whereas for $P(j')$, the second maximum
for large $j'$ is absent from simulations using the PES-PIP. The
high-$j'$ distributions (green and blue in the right column of Figure
\ref{fig:dist-all-jw}) from simulations using the RKHS-PES mainly
originate from rather high initial $v \geq 20$ and high initial $j
\geq 150$ states. In Figure \ref{fig:dist-all-jw}, the high $j'$
distributions correspond to initial condition $(v = 30, j = 150)$ and
$(v = 21, j = 195)$, regardless of the initial $E_{\rm trans}$. Given
the topographies of the two PESs it is conceivable that these
differences arise due to the different angular anisotropies of the
RKHS-PES and PIP-PES.\\

\begin{figure}
\centering \includegraphics[width=0.9\linewidth]{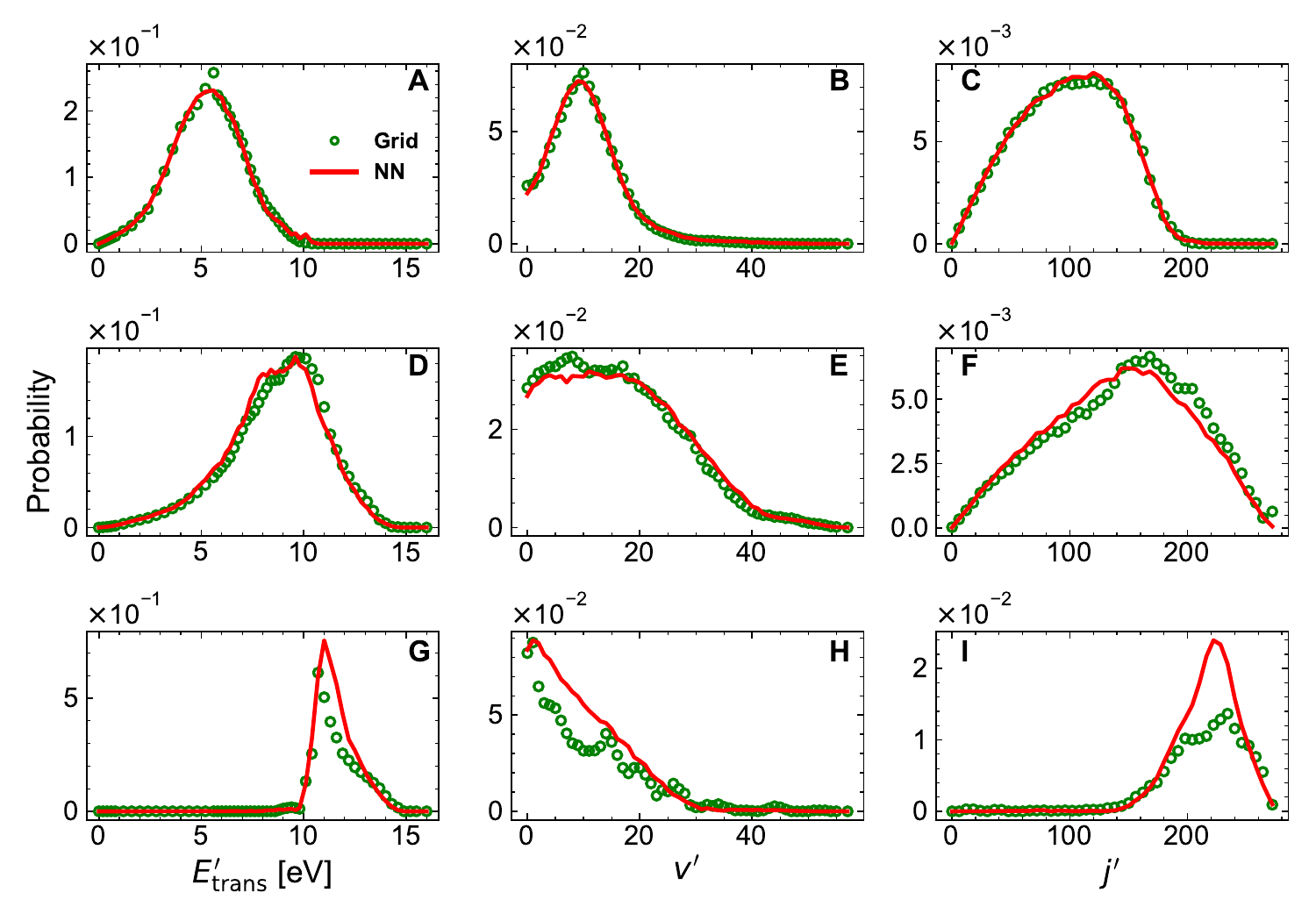}
\caption{STD Model for final state distributions from QCT simulations
  using the RKHS representation for the $^{4}$A$''$ state PES$_{\rm
    Basel}$. Reference amplitudes (Grid) obtained from taking moving
  averages of the raw QCT data in comparison to STD predictions
  (NN). The final $E_{\rm trans}'$, $v'$, and $j'$ are plotted as a
  function of the reaction probability. Three cases for the quality of
  the NN-trained models are distinguished: initial condition for which
  the prediction is best (A to C, highest $R^2 = 1.00$), is closest to
  the average $R^2$ (D to F, $R^2 = 0.97$), and worst (G to I, $R^2 =
  0.64$). The corresponding initial conditions are [$E_{\rm trans} =
    4.5$ eV, $v = 12$, $j = 75$]; [$E_{\rm trans} = 7.0$ eV, $v = 21$,
    $j = 150$]; [$E_{\rm trans} = 4.0$ eV, $v = 27$, $j = 210$] for
  the best, mean and worst cases, respectively.}
    \label{fig:std-unibas}
\end{figure}

\begin{figure}
\centering
\includegraphics[width=0.9\linewidth]{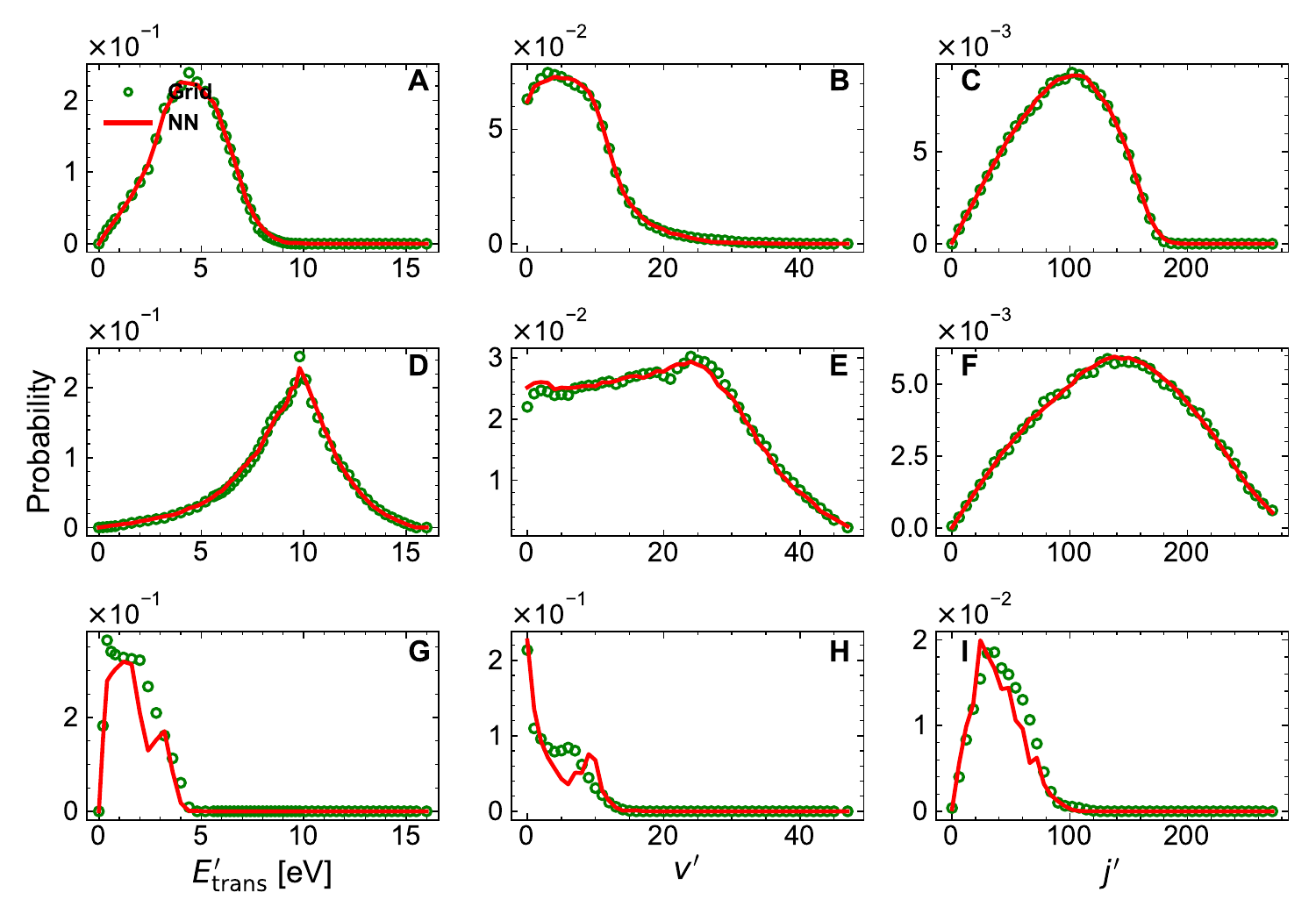}
\caption{STD Model for final state distributions from QCT simulations
  using the PIP-PES for the $^{4}$A$''$ state. Reference amplitudes
  (Grid) obtained from taking moving averages of the raw QCT data in
  comparison to STD predictions (NN). The final $E_{\rm trans}'$,
  $v'$, and $j'$ are plotted as a function of the reaction
  probability. Three cases for the quality of the NN-trained models
  are distinguished: initial condition for which the prediction is
  best (A to C, highest $R^2 = 1.00$), is closest to the average $R^2$
  (D to F, $R^2 = 0.99$), and worst (G to I, $R^2 = 0.92$). The
  corresponding initial conditions are [$E_{\rm trans} = 3.5$ eV, $v =
    4$, $j = 150$]; [$E_{\rm trans} = 6.0$ eV, $v = 12$, $j = 225$];
  [$E_{\rm trans} = 2.5$ eV, $v = 10$, $j = 0$] for the best, mean and
  worst cases, respectively.}
    \label{fig:std-truhlar}
\end{figure}

\noindent
Finally, STD models were trained following the procedures described in
the methods section for both PESs. The performance of both trainings
is reported in Figures \ref{fig:std-unibas} and
\ref{fig:std-truhlar}. For both models the performance on the test set
is reported and comparisons between reference QCT simulations and
trained models are given for the best, the average and the worst
performing final state distributions together with the $R^2-$value
between the NN-model and the results from explicit QCT simulations.\\

\noindent
For the STD model using QCT results based on the RKHS-PES, the overall
$R^2$ on the training and test sets is 0.97 compared with 1.00 and
0.99 for the data based on QCT simulations using the PIP-PES. During
training, the errors drop more rapidly for the data related to the
PIP-PES compared to that from the RKHS-PES. A possible reason is that
the RKHS-PES tends to give broader final state distributions, in
particular for $P(j')$ and $P(v')$, than the PIP-PES out of QCT
simulations for the same initial conditions, see
Figure~\ref{fig:dist-all-jw}. This also suggests that the RKHS-PES is
somewhat more anisotropic than the PIP-PES. Consequently, it requires
a larger number of QCT trajectories to access all the final states for
the RKHS-PES than for the PIP-PES, and training the STD-NN model to
reach the same level of accuracy typically requires more data from
simulations using the RKHS-PES. In this work, the simulation outputs
with final $P(v')<0.02$ were discarded from the training set, which
results in 1900 samples (out of 2360 samples in total) used as the
training set for the RKHS-PES, while the training set of PIP-PES
contains 2000 samples (out of 2314 samples in total).\\

\section{Discussion and Conclusions}
This work investigates thermal rates and final state distributions for
the N($^4$S) + N$_2(^1 \Sigma_{\rm g}^+)$ collision system using a
new, global RKHS-represented PES of high quality. Extensive reference
calculations were carried out at the MRCI-F12/AVTZ-F12 level of theory
to be consistent with previous work on thermal rates and final state
distributions for the [NNO], [OON], and [OOC] collision
systems.\cite{MM.co2:2021,MM.no2:2020,MM.n2o:2020,MM.cno:2018} This
will allow consistent modeling and incorporation of the relevant
microscopic information - such as state-to-state or thermal rates - in
reaction networks encompassing all these species.\cite{boyd:2017} The
MRCI-F12/AVTZ-F12 method was chosen here due to its improved
convergence properties over MRCI+Q/AVTZ but differences in the
relative energetics between the two methods are found to be small,
typically of the order of a few percent.\\

\noindent
The N$_3-$PES was validated for the atom exchange and atomization
reactions using an extensive distribution of initial states. For the
atom exchange reaction, the thermal rates $k(T)$ are in good agreement
with experiment within error bars, whereas for the atomization
reaction, the computed rates are larger than those from experiment by
about one order of magnitude. However, the temperature-dependence is
correctly captured. Atomization rates using the PIP-PES were found to
agree with recent simulations and with experiment, validating the
overall QCT-simulation strategy used here.\\

\noindent
The overestimation of the rate for atomization using the RKHS-PES can
be traced back to differences in the MRCI-F12 and CASPT2 methods and
the basis sets used. First, the N$_2$ molecule is less stable using
the MRCI-based RKHS-represented PES than with the CASPT2-based PIP
representation. To obtain the relevant diatomic potentials $V(r)$, the
third nitrogen atom was placed 500 a$_0$ away from the center of mass
of N$_2$ and the N$_2$ bond length $r$ scanned. Well depths and
dissociation energies for N$_2$ are $D_{\rm e}^{\rm RKHS} = 226.0$
kcal/mol, $D_0^{\rm RKHS} = 222.9$ kcal/mol, compared with $D_{\rm
  e}^{\rm PIP} = 228.4$ kcal/mol, $D_0^{\rm PIP} = 225.1$
kcal/mol. Both $D_0-$values compare favourably with recent experiments
that reported $D_0^{\rm exp} = 225$ kcal/mol,\cite{wang:2024} but the
CASPT2-PIP PES is clearly superior for N$_2-$dissociation by construction because the electronic structure calculations were carried out with this purpose in mind, see Methods Section. The bound
states for N$_2$ were determined using the LEVEL-program.\cite{level}
A second difference between the two PESs is the height of the barrier
to form bent N$_3$ which is lower by 9 kcal/mol for the RKHS-PIP PES
(see Table \ref{tab:struc} and Figure
\ref{fig:N3_PES_2D-r_eq}B). These two effects contribute to finding a
higher thermal rate for the atomization reaction using the
RKHS-PES. Comparable differences of 1 to 3 kcal/mol between MRCISD(Q)
and CASPT2 energies using the AVTZ basis set for triangular N$_3$ and
the TS separating from the N($^4$S) + N$_2(^1 \Sigma_{\rm g}^+)$
asymptote were already reported in earlier work.\cite{wodtke:2005}
Possibilities to further improve the RKHS-PES (e.g. adjust the barrier
to form bent quartet-N$_3$) include PES-morphing using information
from the CASPT2 PES as has recently been done for
He--H$_2^+$,\cite{MM.morph:2024} or to use transfer learning
techniques.\cite{MM.tl:2023}\\

\begin{figure}
\centering
\includegraphics[width=0.9\linewidth]{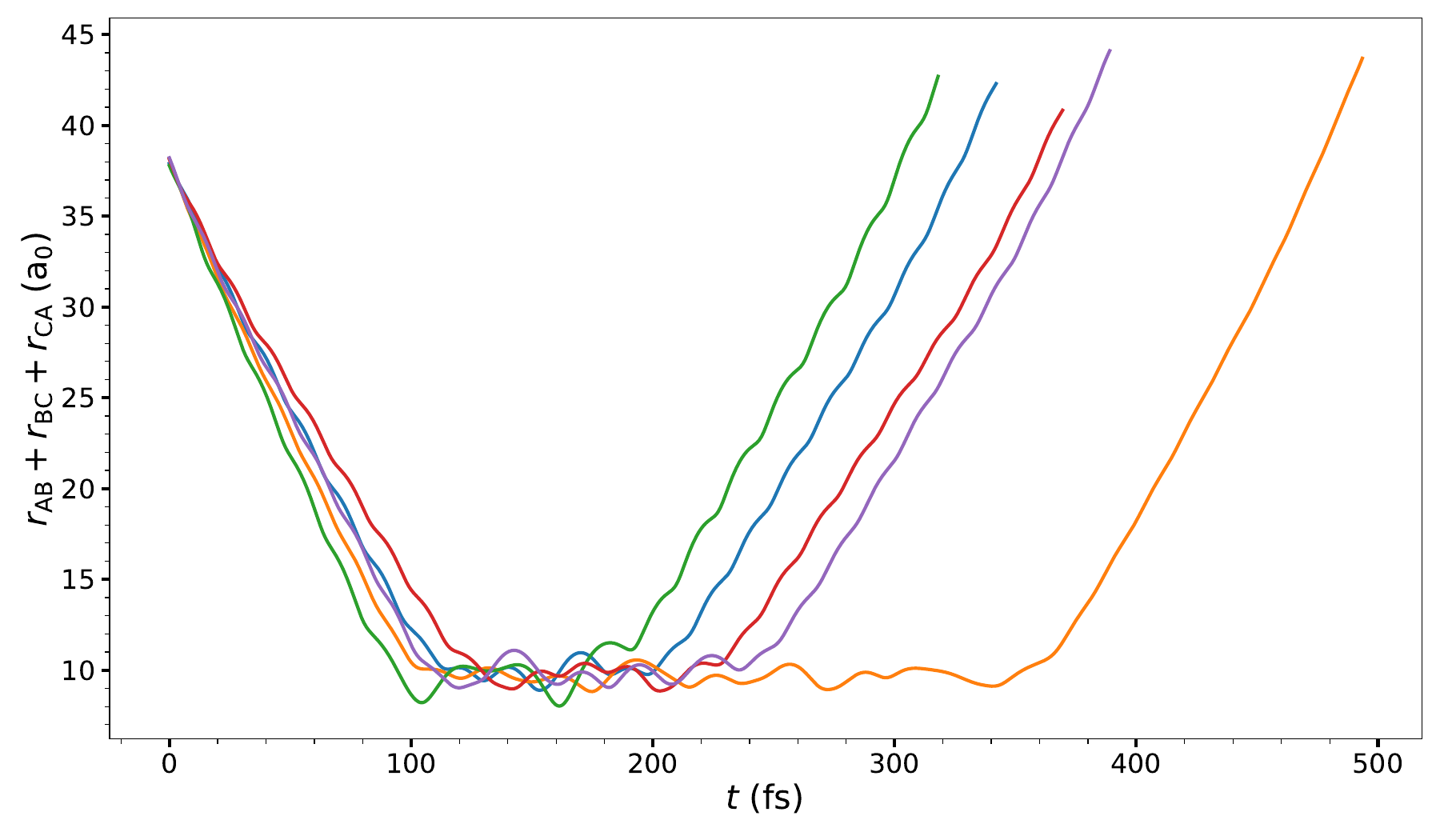}
\caption{Sum $\rho(t) =
  r_{\rm AB}(t)+r_{\rm AC}(t)+r_{\rm BC}(t)$ of atom-atom separations
  as a function of simulation time. N$_3$ in its local minimum energy
  structure for $\rho(t) \simeq 9$ a$_0$, see Table
  \ref{tab:struc}.Typical lifetimes are on the 200 fs time
  scale. Initial conditions are $(E_{\rm trans} = 3.5 {\rm \ eV}, v=1,
  j=2)$ (blue), $(E_{\rm trans} = 4.0 {\rm \ eV}, v=1, j=2)$ (orange),
  $(E_{\rm trans} = 4.5 {\rm \ eV}, v=1, j=2)$ (green), $(E_{\rm
    trans} = 3.0 {\rm \ eV}, v=2, j=2)$ (red), $(E_{\rm trans} = 3.5
  {\rm \ eV}, v=2, j=2)$ (lila).}
    \label{fig:traj}
\end{figure}

\noindent
Finally, it is of interest to consider the probability for stabilizing
bent-N$_3$. First, none of the simulations indicated that c-N$_3$ can
be stabilized for the entire 75 ps for which the QCT simulations were
run. In all cases either atom exchange, atomization or
elastic/inelastic scattering occurred. However, it is possible that
c-N$_3$ forms transiently and a number of atom-exchange trajectories
were analyzed to this end, see Figure \ref{fig:traj}.\\

\noindent
In summary, QCT simulations using the RKHS-based PES for the N($^4$S)
+ N$_2(^1 \Sigma_{\rm g}^+)$ collision system yielded thermal rates
for the atom exchange reaction in favourable agreement with
experiments. The temperature-dependence for the atomization reaction
is consistent with experiments but overestimates absolute values by
one order of magnitude. However, it is straightforward to account for
this in subsequent reaction network simulations. For a comprehensive
description of the state-dependent rates, STD models for both PESs
were determined from an NN-based approach. The performance of both STD
models is on par with previous
systems.\cite{MM.sts:2019,MM.nn:2020,MM.nn:2021,MM.std:2022,MM.std2:2022}
The combination of high-level electronic structure calculations,
machine-learning-based representation of the corresponding PES, QCT
dynamics simulations, and NN-based modeling of the entire initial and
final state space provides a meaningful way to describe the reaction
dynamics of atom + diatom reactions.\\

\section*{Data Availability}
The RKHS-PES, the two STD models, and one video for the atom exchange
reaction are available at \url{https://github.com/MMunibas/N3}.

\section*{Acknowledgment}
We thank the United State Department of the Air Force (AFOSR), the
Swiss National Science Foundation (grants 200021{\_}117810,
200021{\_}215088, and the NCCR MUST) (to MM), and the University of
Basel for supporting this work. Valuable discussions with
Proff. D. Koner, G. Schatz and H. Guo are acknowledged.

\bibliography{ref}

\clearpage

\renewcommand{\thetable}{S\arabic{table}}
\renewcommand{\thefigure}{S\arabic{figure}}
\renewcommand{\thesection}{S\arabic{section}}
\renewcommand{\d}{\text{d}}
\setcounter{figure}{0}  
\setcounter{section}{0}  
\setcounter{table}{0}

\newpage

\noindent
{\bf SUPPORTING INFORMATION: High-Energy Reaction Dynamics of N$_{3}$}

\subsection{The Potential Energy Surfaces}

\begin{figure}
    \centering
    \includegraphics[width=1.0\linewidth]{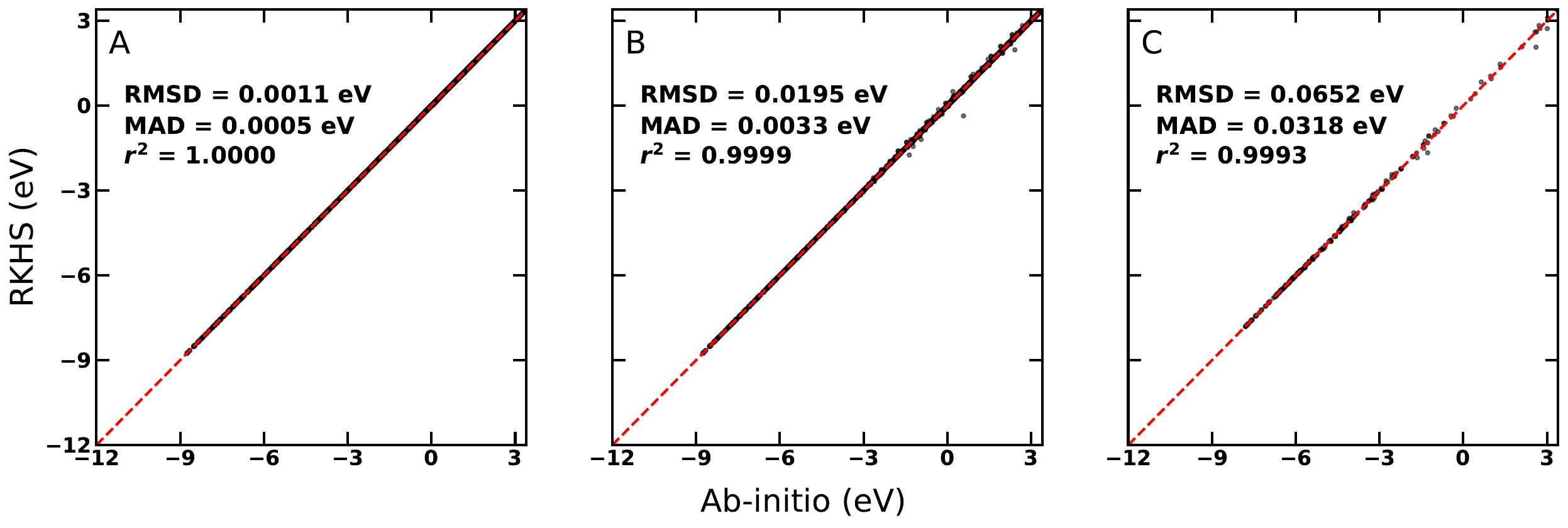} 
    \caption{Correlation between the RKHS energies for on-grid data on
      the single channel PES (panel A), for on-grid data on the 3D
      channel PES (panel B), and for the off-grid validation data in
      comparison to the \textit{ab initio} energies (panel C),
      respectively. The statistical measures (root mean squared
      difference - RMSD, mean absolute deviation - MAD, and $r^2$) for
      each case are given in the panels. The RMSEs/MADs are 0.03/0.01
      kcal/mol, 0.45/0.08 kcal/mol, and 1.50/0.73 kcal/mol from left
      to right.}
    \label{sifig:corr-all}
\end{figure}

\begin{figure}
    \centering
    \includegraphics[width=0.9\linewidth]{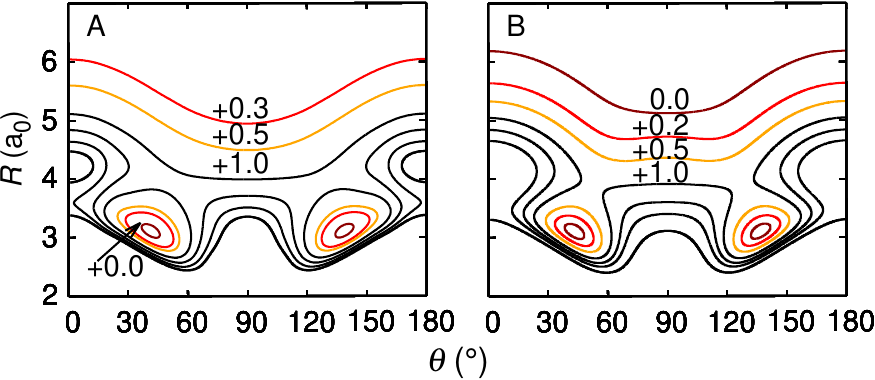}
    \caption{2D contour representations of $V(R,\theta)$ at $r_{\rm
        NN} = 2.5$ a$_0$ for the RKHS-PES (panel A) and for the
      PIP-PES (panel B).\cite{varga:2021} Isocontours are labelled by
      energy difference in eV with respect to the global minimum of
      the specific cut of the PES. Contours at higher energies than
      the +1.0 eV line are at +1.5, +2.0, and +2.5 eV, respectively.}
    \label{sifig:pes-rtheta}
\end{figure}

\begin{figure}
    \centering
    \includegraphics[width=\textwidth]{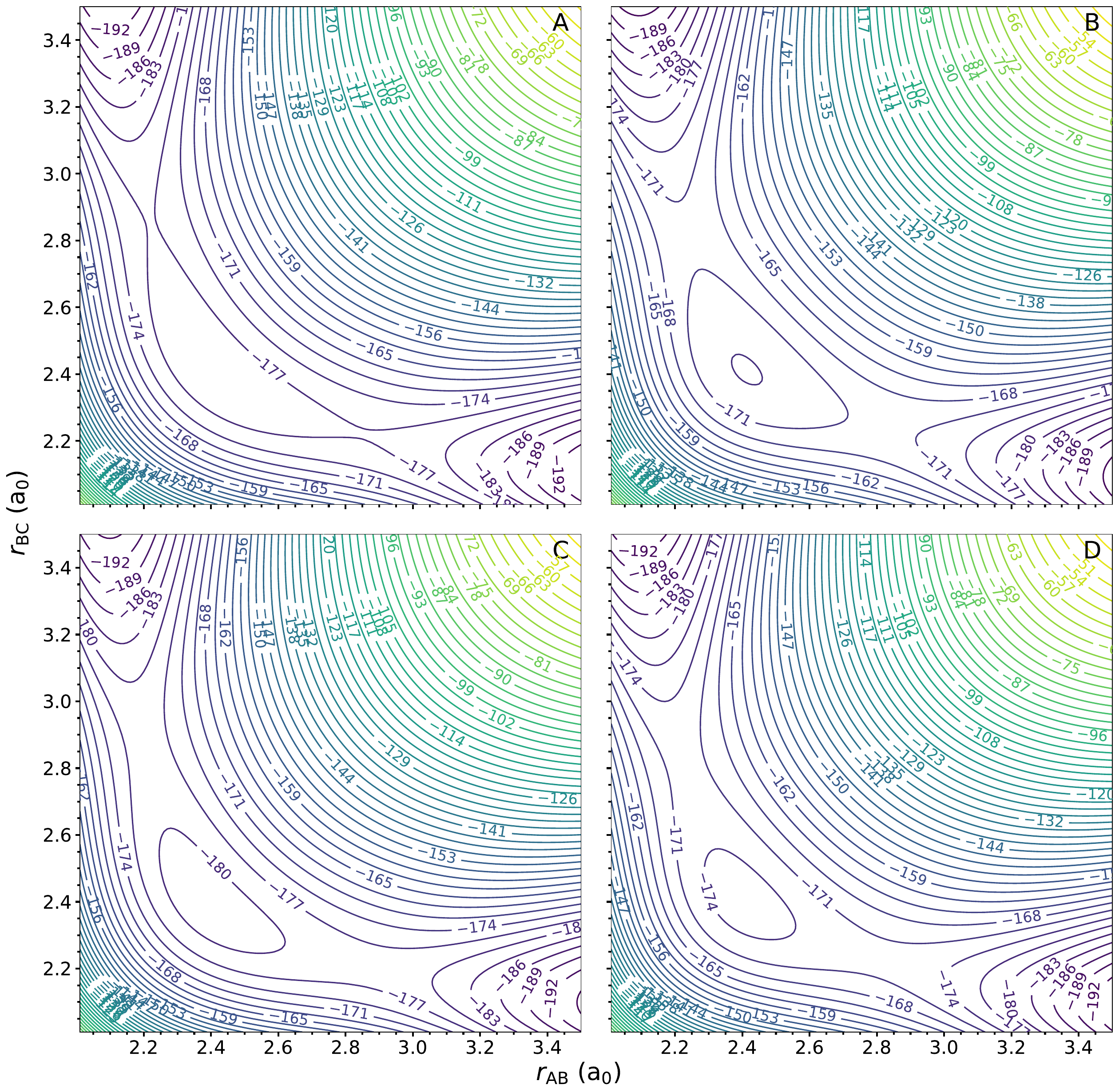}
    \caption{Comparison between the RKHS-PES (left column) and the
      PIP-PES (right column) for N$_{\rm A}$-N$_{\rm B}$-N$_{\rm C}$
      bond angles $110^{\circ}$ (top panels) and $116^{\circ}$ (bottom
      panels).  Panels A to D show isocontours of the RKHS-PES at bond
      angle $110^{\circ}$; PIP-PES at bond angle $110^{\circ}$;
      RKHS-PES at bond angle $116^{\circ}$; PIP-PES at bond angle
      $116^{\circ}$.  $r_{\rm AB}$ and $r_{\rm BC}$ are separations
      N$_{\rm A}$--N$_{\rm B}$ and N$_{\rm B}$--N$_{\rm C}$,
      respectively. Energies are in kcal/mol.}
    \label{sifig:pes-r1r2}
\end{figure}

\end{document}